# Threshold Regression for Survival Analysis: Modeling Event Times by a Stochastic Process Reaching a Boundary

Mei-Ling Ting Lee and G. A. Whitmore

*Abstract.* Many researchers have investigated first hitting times as models for survival data. First hitting times arise naturally in many types of stochastic processes, ranging from Wiener processes to Markov chains. In a survival context, the state of the underlying process represents the strength of an item or the health of an individual. The item fails or the individual experiences a clinical endpoint when the process reaches an adverse threshold state for the first time. The time scale can be calendar time or some other operational measure of degradation or disease progression. In many applications, the process is latent (i.e., unobservable). *Threshold regression* refers to first-hitting-time models with regression structures that accommodate covariate data. The parameters of the process, threshold state and time scale may depend on the covariates. This paper reviews aspects of this topic and discusses fruitful avenues for future research.

## 1. INTRODUCTION

Many types of lifetime, duration or time-to-event data may be interpreted as *first hitting times* (FHT's) of a boundary or threshold state by sample paths of a stochastic process, which may be latent or observable. FHT models have a long history of application in diverse fields, including medicine, environmental science, engineering, business, economics and sociology. FHT models may describe the length of a hospital stay, the survival time of a transplant patient, the onset time for a cancer induced by occupational exposure, the failure time of an engineering system, the depletion time of an inventory, the survival time of a new business, the transition time for a stock price change and the length of a marriage. Relevant articles, covering theory and application, appear in both the lifetime data and reliability literatures. These FHT models are gradually becoming widely adopted because of their conceptual appeal, realism and applicability. Recently, interest in them has deepened and spread, and exciting new areas of application are being encountered. The potential applications require new conceptual viewpoints, theoretical advances, analytical techniques and methodological extensions to which the discussion returns later.

*Mei-Ling Ting Lee is Professor and Chair, Division of Biostatistics, School of Public Health, Ohio State University, Columbus, Ohio 43210, USA e-mail: meilinglee@sph.osu.edu. G. A. Whitmore is Professor, Management Science, Desautels Faculty of Management, McGill University, Montreal, Quebec, Canada H3A 165 e-mail: george.whitmore@mcgill.ca.*

To make FHT models truly valuable in applications, they must be capable of extension to include regression structures. Regression structures allow the effects of covariates to explain the inherent dispersion of the data, thereby taking account of variability and sharpening inferences. Regression structures also provide scientific insights into potential causal roles of covariates in the underlying processes, boundary sets and time scales. This article reviews aspects of FHT models and is concerned especially with regression structures for FHT models, which will be referred to as *threshold regression*, or TR for short. The word "threshold" refers to the fact that the FHT is triggered by the underlying process reaching a threshold state within a boundary set, as described in more detail in the next section.

## 2. THE FIRST-HITTING-TIME (FHT) MODEL

A first-hitting-time (FHT) model has two basic components: (1) a *parent stochastic process* $\{X(t), t \in \mathcal{T}, x \in \mathcal{X}\}$ with initial value $X(0) = x_0$, where $\mathcal{T}$ is the time space and $\mathcal{X}$ is the state space of the process and (2) a *boundary set* $\mathcal{B}$, where $\mathcal{B} \subset \mathcal{X}$. We shall refer to the boundary set $\mathcal{B}$ as a boundary, barrier or threshold, depending on which is most descriptive or conventional in the context. The process $\{X(t)\}$ may have a variety of properties, such as one or many dimensions, the Markov property, continuous or discrete states, and monotonic sample paths. Whether the sample path of the parent process is observable or latent (unobservable) is an important distinguishing characteristic of the FHT model. Latent processes are the most common by far. The boundary set $\mathcal{B}$ may also have different features as will be illustrated in later examples.

Taking the initial value $X(0) = x_0$ of the process to lie outside the boundary set $\mathcal{B}$, the first hitting time of $\mathcal{B}$ is the random variable $S$ defined as

$$(1) \qquad S = \inf\{t : X(t) \in \mathcal{B}\}.$$

Thus, the first hitting time is the time when the stochastic process first encounters set $\mathcal{B}$. We refer to the state first encountered in the boundary set by the process, that is, $X(S) \in \mathcal{B}$, as the *threshold state*. The boundary set defines a stopping condition for the process and, therefore, the FHT is usually a stopping time in the formal sense of that term in stochastic process theory. Note that when the parent process is latent, there is no direct way of observing the FHT event in the state space of the process.

In some versions of the FHT model, there is no guarantee that process $\{X(t)\}$ will reach the boundary set $\mathcal{B}$, so $P(S < \infty) < 1$. We will let $S = \infty$ denote the absence of a finite hitting time with $P(S = \infty) = 1 - P(S < \infty)$. The later discussion will show situations where this condition is plausible and a desirable model feature. The basic FHT model (1) assumes that $\mathcal{B}$ is fixed in time. In some applications, however, it varies with time, that is, $\mathcal{B}(t)$. This variation may be deterministic or follow a stochastic process.

An exhaustive review of the first-hitting-time literature is impossible within the confines of a single article. Eaton and Whitmore (1977) discuss FHT's as a general model for hospital stay. Aalen and Gjessing (2001) provide an excellent overview of much of this subject. Likewise, Lawless (2003) gives a complete and compact summary of theory, models and methods (see Section 11.5, pages 518–523). Lee and Whitmore (2004) also provide an overview of first-hitting-time models for survival and time-to-event data. We will make numerous references to selected work as we proceed. There is a huge literature dealing with theoretical and mathematical aspects of FHT models that we will not attempt to review or incorporate. We also will not cover random growth curve models, such as those of Carey and Koenig (1991) and Lu and Meeker (1993), which have an FHT interpretation but where the only randomness at the level of the individual parent process is confined to a noise factor. The large literature on linear and nonlinear regression methods for survival data and reliability where the underlying models have no central FHT motivation (such as accelerated failure time and proportional hazards models) is also not covered.

## 3. EXAMPLES OF FIRST-HITTING-TIME MODELS

The parent stochastic processes may take many forms, from Wiener processes to Markov chains. Likewise, the nature of the boundary state may vary widely—for example, a fixed threshold in a Wiener process or an absorbing state in a Markov chain. As the preceding description of a first-hitting-time model is quite abstract, we now list a few basic examples to illustrate the variety encountered in applications.

1. *Bernoulli process and negative binomial first hitting time.* The number of trials $S$ required to reach the $m$th success in a Bernoulli process $\{B_t, t = 1, 2, \ldots\}$ has a negative binomial distribution with parameters $m$ and $p$, where $p$ is the success probability on each trial. To give this setup our standard representation, we consider a parent process $\{X_t, t = 0, 1, 2, \ldots\}$ with initial value $X_0 = x_0 = m$ and let $X_t = x_0 - B_t, t = 1, 2, \ldots$, where $\{B_t\}$ is the preceding Bernoulli process. The first hitting time is the first Bernoulli trial $t = S$ for which $X_t$ equals 0. The number of rocket launches required to get $m$ satellites in orbit is a simple example of this FHT model.

2. *Poisson process and Erlang first hitting time.* The time $S$ until the occurrence of the $m$th event in a Poisson process $\{N(t), t \geq 0\}$ with rate parameter $\lambda$ has an Erlang distribution with parameters $m$ and $\lambda$. Again, to give this setup our standard representation, we consider a parent process $\{X(t), t \geq 0\}$ with initial value $X(0) = x_0 = m$ and let $X(t) = x_0 - N(t)$, where $\{N(t), t \geq 0\}$ is the preceding Poisson process. The first hitting time is the earliest time $t = S$ when $X(t) = 0$. This FHT model is illustrated by the time to failure of an engineering system consisting of $m$ components in parallel, having identical and independent exponential lifetimes, that are placed in service successively as failures occur.

3. *Wiener process and inverse Gaussian first hitting time.* Consider a Wiener process $\{X(t), t \geq 0\}$ with mean parameter $\mu$, variance parameter $\sigma^2$, and initial value $X(0) = x_0 > 0$. The time $S$ required for the process to reach the zero level for the first time has an inverse Gaussian distribution if the process mean parameter $\mu$ is negative so the process tends to drift toward the zero level. Lancaster (1972) describes the duration of an industrial strike using this model. In this context, $\{X(t)\}$ represents the distance between the positions of management and labor at time $t$ after the start of the strike. The initial separation of the parties is $X(0) = x_0 > 0$. The strike ends when the process first encounters the zero level where the parties agree and settle. In addition to Lancaster (1972), refer to applications described in Whitmore and Neufeldt (1970), Whitmore (1975, 1979, 1983, 1986, 1995), Doksum and Hóyland (1992), Doksum and Normand (1995), Lu (1995), Whitmore and Schenkelberg (1997) and Horrocks and Thompson (2004), to name a few. Onar and Padgett (2000) and Padgett and Tomlinson (2004) extend the Wiener diffusion model to an accelerated testing context. Pettit and Young (1999) set the model in a Bayesian context. As illustrated later, the inverse Gaussian distribution has a closed-form probability density function (p.d.f.) and a computationally simple cumulative distribution function (c.d.f.). Their formulas vary slightly depending on whether the parent process is defined as rising or falling to hit the relevant boundary.

4. *Gamma process and inverse gamma first hitting time.* Consider a parent process $\{X(t), t \geq 0\}$ with initial value $X(0) = x_0 > 0$. Let $X(t) = x_0 - Z(t)$ where $\{Z(t), t \geq 0\}$ is a gamma process with scale parameter $\beta$, shape parameter $\alpha$ and $Z(0) = 0$. The first hitting time of the zero level in the parent process $(X = 0)$ has an inverse gamma c.d.f., defined by the identity $P(S > t) = P(Z(t) < x_0)$. The identity follows from the fact that a gamma process has monotonic (nondecreasing) sample paths. The availability of computational routines for the gamma c.d.f. allows the c.d.f. of $S$ to be computed readily. Singpurwalla (1995) and Lawless and Crowder (2004) consider the gamma process as a model for degradation. Park and Padgett (2005) consider both geometric Brownian motion and gamma processes in an accelerated degradation model.

5. *Ornstein–Uhlenbeck process and Ricciardi–Sato first hitting time.* The Ornstein–Uhlenbeck (OU) process is a variant of a Wiener process that is mean-reverting in that it tends to drift back toward a fixed equilibrium level and thus has a property of homeostasis. Aalen and Gjessing (2004) propose the first-hitting-time distribution for such a process as a survival model and derive pertinent results. They point out that the form of the FHT distribution is found in Ricciardi and Sato (1988), who have studied it extensively.

6. *Markov chain and absorbing state first hitting time.* Markov chains $\{X_t, t = 0, 1, 2, \ldots\}$ are an important type of parent process. The state space $\mathcal{X}$ consists of the possible states of the chain. The time space $\mathcal{T}$ is the transition steps for the chain. The first hitting time is the minimum number of steps required to move from an initial state $X_0 = x_0$ to a set of boundary states $\mathcal{B}$. The FHT distribution depends on the transition matrix of the chain in a precise mathematical manner. As a case example, a Markov chain can model product brand switching in the field of consumer behavior. An FHT of interest might be the number of purchases that will be made in the product category before a consumer who currently uses brand $x_0$ will switch to another brand, say, $b$. In this case, $\mathcal{B}$ is the singleton set $\{b\}$.

7. *Semi-Markov processes and their first hitting times.* All of the preceding examples are special cases of Markov processes. A semi-Markov process $\{X(t), t \geq 0\}$ extends the Markov chain model by including the random time that the process resides in each state. Although the Markov property is generally lost by this extension, the model remains of great practical value. In a semi-Markov model, the first hitting time represents the time that the process resides in the initial and subsequent states before it first enters one of the states that define set $\mathcal{B}$. There are many important examples of this multi-state model, some more complicated than others. The *two-stage clonal conversion* model for cancer provides a case example. This is a chemical carcinogenesis model that, in its basic form, postulates a pool of (stem) cells that are susceptible to malignant transformation. The cells proceed through an initiation stage and then through malignant conversion according to a two-stage Markov process with fixed transition rates. Once a malignant cell is generated, a second statistical model describes progression through stages of active cancer to death (homeostasis, angiogenesis, metastasis, death). The model has been elaborated in a series of works by Moolgavkar, Luebeck and Anderson (1998) and Luebeck et al. (1999), and others. The model has been applied, for example, in a study of lung cancer risk posed to Chinese tin miners by arsenic, radon and tobacco exposure (Hazelton et al., 2001).

## 4. LATENT FIRST HITTING TIMES AND COMPETING RISKS

Most duration data are gathered under conditions of competing risks in which two or more causes are competing to determine the observed duration. The outcome becomes associated with both a *time* and *mode* of occurrence. For example, a medical first-hitting-time model that describes the time of a subject's death may recognize explicitly that multiple causes are competing to produce death. Both the time and cause of death are recorded for each subject. Refer to Kalbfleisch and Prentice (1980, 2002) and Crowder (2001), for example, for detailed technical discussions of this topic.

FHT models accommodate the competing risk aspect in a natural fashion. In an FHT model with competing risks, the boundary set $\mathcal{B}$ is partitioned into mutually exclusive and exhaustive subsets, say $\mathcal{B}_1, \ldots, \mathcal{B}_C$, associated with FHT causes $c = 1, \ldots, C$. Let $D$ denote the observed cause. Then realization $D = d$ associated with the observed FHT outcome $S$ would be defined as

$$(2) \qquad D = d \quad \text{if } X(S) \in \mathcal{B}_d.$$

The concept of a *latent* FHT offers an interesting vehicle for discussing competing risks. In this framework, an individual is imagined to have latent FHT's $S_1, \ldots, S_C$ for the $C$ competing causes. The FHT $S_c$ is defined as in (1) with $\mathcal{B}_c$ replacing $\mathcal{B}$. The observed cause $d$ and observed FHT $S_d$ are then given by

$$(3) \qquad S_d = \min\{S_c, c = 1, \ldots, C\}.$$

In other words, the observed FHT time and mode are those of the smallest latent FHT.

As a practical matter, latent FHT's, other than the smallest, are generally unobservable, although there may be exceptions. For example, an engineering system may consist of $C$ components and fail whenever one of its components fails. If the system is repairable, then the failure times of all $C$ components may eventually be observed because repair allows the system to continue functioning. It is questionable, however, if a repaired system really continues on the same stochastic trajectory as before the repair. Nevertheless, latent FHT's have value as notional measures of survival increments that might be realized in an idealized world where selected causes of failure are removed. In terms of the definition in (3), an investigator might look at the impact on the observed FHT if some boundary subsets $\mathcal{B}_c$, associated with selected causes, are removed from the model formulation. This removal, in effect, eliminates the associated latent FHT's from (3). This exercise therefore simulates the elimination of some causes of death (in medicine) or failure (in engineering). An FHT model offers a clear conceptual structure for the status of an individual with respect to modes of failure other than the one observed. This structure is captured by the "distance" that the threshold state $X(S_d)$ lies from the boundary set $\mathcal{B}_c$ for any cause $c$ that differs from the observed cause $d$. The difference $S_c - S_d$ for $c \neq d$ reflects the same distance on the survival time scale.

FHT models offer the possibility of inferences about this distance.

To illustrate an FHT competing risks model by a concrete medical example, consider a multidimensional Wiener process of $C$ dimensions with a boundary $\mathcal{B}_c$ in each dimension. Each of the $C$ dimensions defines a different cause of death $c \in \{1, 2, \ldots, C\}$. In such a model, one may make inferences about secondary medical conditions that are not the primary cause of death. For example, in studies of occupational exposure to diesel exhaust, workers may be found to have increased risks of death from lung disease, cardiovascular disease and other causes. It is desirable in such a context to have an FHT model that is capable of considering different causes of death simultaneously. A worker dying of lung cancer (the primary cause of death) may have advanced cardiovascular disease, both of which are aggravated by exposure to an occupational hazard such as diesel exhaust. Then, an investigator's interest may lie in making inferences about the worker's cardiovascular disease status at the time of death from lung cancer. We note that if the underlying multidimensional Wiener process is correlated, then the latent survival times for different causes of death will be dependent.

## 5. CURE RATES

We mentioned at the outset that some FHT models may offer a positive probability of no FHT taking place in finite time. Thus, for example, a medical treatment may offer a cure, some animals in a population may be immune to infection, some stock prices may never reach $100, and some marriages may never end in divorce. The fact that $P(S = \infty) > 0$ in some FHT models is closely related to competing risks. Generally, if the FHT model takes account of *all* competing risks, then eventual failure from some cause is assured. If, however, the FHT model takes account of only one or a few competing risks, then there is a positive probability that the FHT will be infinite to accommodate those individuals who are not susceptible to the limited array of causes of failure that are considered in the model. To illustrate the natural way in which FHT models take account of a cure rate, consider a Wiener diffusion model with a fixed boundary at zero (the time axis). If the drift of the process is away from the boundary, that is, $\mu > 0$, then a finite FHT is not assured and, in particular, $P(S < \infty) = \exp(-2x_0\mu/\sigma^2)$. Likewise, a gamma process with a cure rate might be defined as

$$(4) \quad X(t) = \begin{cases} x_0, & \text{with probability } 1-p, \\ x_0 - Z(t), & \text{with probability } p. \end{cases}$$

Here the parameter $p$ is a susceptibility fraction, with $0 \le p \le 1$. As an example of this last model, a subject may have a malignant or benign form of a disease with probabilities $p$ and $1-p$, respectively. The malignant form progresses monotonically toward a medical endpoint (e.g., death).

## 6. COVARIATES AND LINK FUNCTIONS FOR THRESHOLD REGRESSION

The parent process $\{X(t)\}$ and boundary set $\mathcal{B}$ of the FHT model will both generally have parameters that depend on covariates that vary across individuals. To illustrate, consider the Wiener process model in Example 3. The Wiener process has mean parameter $\mu$ and variance parameter $\sigma^2$ and the boundary set has parameter $x_0$, the initial process level. In threshold regression, these parameters will be connected to linear combinations of covariates using suitable regression link functions, as illustrated below for some parameter, say $\theta$,

$$(5) \quad g_\theta(\theta_i) = \mathbf{z}_i \boldsymbol{\beta}.$$

Here $g_\theta$ is the link function, the parameter $\theta_i$ is the value of the parameter $\theta$ for individual $i$, $\mathbf{z}_i = (1, z_{i1}, \ldots, z_{ik})$ is the covariate vector of individual $i$ (with a leading unit to include an intercept term) and $\boldsymbol{\beta}$ is the associated vector of regression coefficients. The mathematical form of the link function must be suited to the application. Generally, it will be chosen to map the parameter space into the real line. For example, a variance parameter such as $\sigma^2$ may employ a logarithmic link function, that is, $\ln(\sigma^2) = \mathbf{z}\boldsymbol{\beta}$. Likewise, the list of covariates and their mathematical forms in the regression function $\mathbf{z}\boldsymbol{\beta}$ must be chosen appropriately, as is the case in a conventional regression analysis.

Previous work that has considered regression structures for FHT models includes Whitmore (1983), Whitmore, Crowder and Lawless (1998), Lee, DeGruttola and Schoenfeld (2000) and Lee et al. (2004). To illustrate one of these applications, Lee, DeGruttola and Schoenfeld (2000) use a bivariate Wiener diffusion process as the basis of a threshold regression model for the study of progression to death in AIDS, with CD4 cell count serving as a marker process (marker processes are discussed in a later section). The initial health status and mean parameter of the parent process are made to depend on baseline covariates and treatment variables through log-linear and identity link functions, respectively. The mean and variance parameters of the marker process are also given a regression structure, with identity and log-linear link functions. Finally, the correlation parameter for the parent and marker processes uses a correlation transform as a link function.

Threshold regression raises some new issues for estimation and inference in FHT models. Where FHT models are estimated only from censored survival data, parameter estimators may exhibit significant multicollinearity, especially with highly parameterized regression functions. This fact does not reflect any deficiency of the FHT model but, rather, reflects the limited information content of sample data in a rich modeling context. Reparameterization of the model can assist with computational problems that may arise from this multicollinearity but generally the condition is not sufficiently severe to prevent estimates from being computed. The impact is primarily felt in the interpretability of the estimation results. As with conventional regression, where regression effects are highly collinear, it will be difficult to attribute the effect to a particular model component. For example, in threshold regression based on censored inverse Gaussian survival data, estimates of covariate effects of the initial value $x_0$ and mean parameter $\mu$ may be collinear because the mean survival time depends on their ratio $x_0/|\mu|$. Thus, the high correlation of their sampling errors can only be mitigated by having fine details for the dispersion pattern of the survival data. Censoring or small sample sizes may mask those fine details and thus make estimation more difficult.

## 7. RUNNING TIME VERSUS CALENDAR TIME

In many applications of threshold regression, the natural time scale of the parent process is not calendar or clock time. For example, a mechanical component may wear according to the amount of its usage or liver disease may progress according to an individual's cumulative consumption of alcohol. Mathematical research on different time scales has been carried out by many researchers. Cox and Oakes (1984, Section 1.2, pages 3–4) pointed out that "often the 'scale' for measuring time is clock time, although other possibilities certainly arise, such as the use of operating time of a system, mileage of a car, or some measure of cumulative load encountered." These accumulation measures are increasing with calendar time and thus are alternative progression scales for the stochastic process. Such measures are given a variety of names, depending on the context, such as *operational time*, *disease progression*, or *running time*. We shall mainly use the last name here. If $r(t)$ denotes the transformation of calendar time $t$ to running time $r$, with $r(0) = 0$, and $\{X(r)\}$ is the parent process defined in terms of running time $r$, then the resulting process expressed in terms of calendar time is the *subordinated process* $X^*(t) = X[r(t)]$, where the asterisk identifies the subordinated process. Adaptations of FHT models to running time scales may be done in a variety of ways, as we describe below. The variety includes both random and nonrandom transformations. We note that $r(t)$ must be a monotonic transformation but its monotonicity need not be strictly increasing. Interesting effects arise, for example, where $r(t)$ is a function with jump discontinuities.

1. Some applications require a monotonic mathematical transformation of the time scale. In these cases, $r(t)$ is a deterministic function of calendar time $t$. A typical example from an engineering application is the strictly monotonic transformation $r = 1 - \exp(-\lambda t^\gamma)$ with $\lambda > 0$ and $\gamma > 0$. See, for example, Carey and Koenig (1991), Whitmore (1995) and Whitmore and Schenkelberg (1997). The mathematical transformation may depend on covariates, as in Bagdonavičius and Nikulin (2001), where the running time scale forms part of an accelerated life model.
2. Running time may also enter an FHT model using a stochastic process for subordination. In this context, the parent process $\{X(r)\}$ is directed by a second stochastic process $\{R(t)\}$ having monotonic sample paths. In this context, we refer to $\{R(t)\}$ as the *directing process* and the subordinated process takes the form $\{X^*(t)\} = \{X[R(t)]\}$. Unlike a monotonic mathematical transformation, subordination with a stochastic process gives the transformation random properties that can greatly enrich the model. Lee and Whitmore (1993) examine the connection between subordinated stochastic processes and running time. As a specific example of a subordinated process, one can consider a Poisson parent process that is directed by a gamma process (which has monotonic sample paths). The result is a clustering Poisson process (Hougaard, Lee and Whitmore, 1997) in which an FHT can be triggered by the occurrence of a cluster of Poisson events.

3. The running time scale may be a combination of different accumulation measures. For example, Oakes (1995) and Kordonsky and Gertsbakh (1997) look at multiple running time scales in survival data analysis. Duchesne and Lawless (2000) and Duchesne and Rosenthal (2003) describe various advances in running time models for survival data. The concept of collapsible time within the context of accelerated failure time models is central to this earlier work. The basic idea appears in various forms. For example, a *composite running time* might be defined by

$$(6) \qquad r(t) = \sum_{j=1}^{J} \alpha_j r_j(t),$$

where the $r_j(t)$ are different accumulation measures that can advance degradation or disease progression and the $\alpha_j$ are positive parameters that weight the contributions of the different measures. One of the measures, say $r_1(t)$, may be calendar time itself so $r_1(t) = t$. One $\alpha_j$ parameter will need to be set to unity to give a well-defined scale. Typically, in this setup, composite running time has a fixed mathematical form for any given individual case but individuals will have different scales because the $r_j(t)$ vary randomly among individuals. As a simple example of a composite running time, consider the mechanical aging of a motor vehicle which may be related to both the passage of calendar time $r_1(t) = t$ and accumulated mileage $r_2(t)$. In this case, (6) has two components, as follows: $r(t) = t + \alpha r_2(t)$. Notice $\alpha_1$ is set to 1 and $\alpha_2 = \alpha$.

A practical case of the last kind of running time is illustrated in Lee et al. (2004) where railroad workers are employed in different types of jobs, indexed by $j = 1, \ldots, J$, which have differential exposures to diesel exhaust, an occupational risk. The running time (6) here is defined as a weighted sum of different exposure intervals. The quantity $r_j(t)$ is the time spent by the worker in job type $j$ during time interval $[0,t]$. The $\alpha_j$ are positive weights that determine the rates at which the running time advances per unit of calendar time spent in the different job types. One $\alpha_j$ is set to unity as a numeraire, say $\alpha_J = 1$. The $r_j(t)$ also must satisfy the accounting constraint $\sum_{j=1}^{J} r_j(t) = t$. Observe that (6) is a deterministic transformation for any given set of exposure intervals $r_j(t)$ but that these intervals vary randomly from one worker to another according to their individual work histories.

Some processes may be defined in terms of running time from the outset. For example, in a Bernoulli process or a Markov chain, the progress parameter represents the sequence of trials or steps of the processes, respectively. These parameters may already be seen as reflecting a kind of running time. The mapping of calendar time to this running time, as represented by $r(t) = r$, is already implicit in the discrete progress parameter of the process.

The running time scale $r(t)$ is included in the FHT model in order to make the model a more valid representation of reality. With a correct specification of running time, one would expect health status or component strength to decline steadily and predictably against the scale that measures the accumulating "wear and tear" of running time. In other words, $X(r)$ would retain very little or no inherent variability if $r(t)$ could be chosen carefully. This situation describes an ideal that is unattainable in most practical applications of FHT models but is a target of model building.

## 8. INCORPORATING MARKER PROCESSES IN THRESHOLD REGRESSION

A *marker process* refers to an external process that covaries with the parent process. It assists in tracking progress of the parent process if the parent process is latent or only infrequently observed. In this way, the marker process forms a basis for predictive inference about the status of the parent process and its progress toward an FHT. Marker processes may also be of scientific interest in their own right. As markers of the parent process, they offer potential insights into the causal forces that are generating the movements of the parent process. Examples of marker processes include CD4 cell count for AIDS, blood pressure for cardiovascular disease, personal medical cost for health status, input drive current for a laser, and ambient temperature for equipment.

The basic analytical framework for a marker process conceives of a bivariate stochastic process $\{X(r), Y(r)\}$ where the parent process $\{X(r)\}$ is one component process and the marker process $\{Y(r)\}$ is the other. Both are assumed to be one-dimensional for convenience of exposition. They are also both defined on the running time scale $r$. We discuss the implications of this last point shortly. Whitmore, Crowder and Lawless (1998) look at failure inference based on a bivariate Wiener model in which failure is governed by the FHT of a latent degradation process while auxiliary readings are available from a correlated marker process. As noted earlier, Lee, DeGruttola and Schoenfeld (2000) apply this bivariate marker model to CD4 cell counts in the context of AIDS survival.

An application may offer a variety of marker processes, say, $\{Y_k(r), k = 1, \ldots, K\}$, that may be of potential scientific value. They can be studied separately or combined into a *composite marker process*. For marker processes that involve measurements, the following additive form for the composite marker might be appropriate:

$$(7) \qquad Y(r) = \gamma_0 + \sum_{k=1}^{K} \gamma_k Y_k(r).$$

The concept of a composite marker was first proposed by Whitmore, Crowder and Lawless (1998) in an engineering context. The aim in constructing the composite marker process is to find that linear combination of the $K$ candidate markers that has the largest predictive correlation or association with the parent process. The $\gamma_k$ parameters define the linear combination and these generally must be estimated. The approach is reminiscent of regression analysis with the composite marker serving as a regression function to predict the parent process $\{X(r)\}$. Here $\gamma_0$ serves as the intercept term of the regression relationship. If the composite marker can mimic the parent process perfectly, then $\{X(r)\}$ and $\{Y(r)\}$ will be perfectly correlated. An exact model for the preceding setup is a $(k+1)$-variate Wiener diffusion process in which the parent process is one component and the $k$ markers are the remaining components. The conditional process $\{X(r)|Y_1(r) = y_1(r), \ldots, Y_K(r) = y_K(r)\}$ then defines an exact linear regression structure. Where one is dealing with a parent process or marker processes that are not measurement processes, such as Markov chains, the concept of a composite marker process must be redefined in a suitable manner.

The FHT modeling framework has evolved in the literature to encompass three major components as shown in Figure 1, namely, an FHT model (parent process and boundary set) that defines the relevant endpoint, a running time scale and a marker process. The threshold regression (TR) structure stands in the background of the schematic in Figure 1 and allows the parameters of the various components in the figure to depend on baseline and other covariates. Although the schematic shows only one marker, it is clear that there may be many. The framework in Figure 1 has several noteworthy features. For example, if a marker process has monotonic sample paths, it may serve either as a marker or as a running time $r(t)$, as may be deemed appropriate by the investigator. The framework reminds us that a marker process $\{Y(r)\}$ should be defined on the running time scale $r$ when its correlation with the parent process $\{X(r)\}$ is being considered. Thus, for example, if $r(t)$ measures an individual's cumulative exposure to a potential carcinogen at time $t$ and the marker $y$ is a serum measurement for the individual on a cancer-specific antigen at time $t$, then the serum reading $y$ should be recorded as a function of cumulative exposure $r$. In other words, the progress parameter of the serum marker process is cumulative exposure $r$, not calendar time $t$.

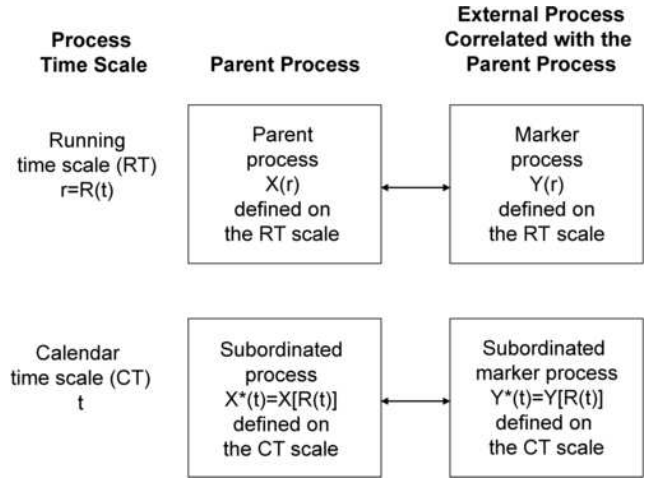

FIG. 1. *This conceptual framework shows the connections between the parent process (often a latent process), running time (RT) and an external marker process that is correlated with the parent process. Time subordination links calendar time (CT) to running time. The threshold regression structure stands in the background and is not displayed explicitly in the figure.*

We have said that the parent process is generally latent. This feature definitely is common in medical applications where inherent health condition cannot be observed (and, indeed, may be deemed unmeasurable). Marker processes are surrogates for health status, especially if they are highly correlated with the underlying medical condition. These markers may range from biomedical measurement processes, such as serum measurements, to more qualitative processes, such as periodic subjective evaluations of health status by a patient or caregiver. In engineering systems, there will be contexts in which wear and tear, for example, can be observed and measured. In many physical settings, however, only surrogates for the system condition are available. For example, the drive current of a laser is a marker for its physical condition but not perfectly correlated with it. Similar comments can be made about social and economic systems. The parent processes that define the FHT (e.g., business failure) may only be imperfectly monitored by a marker process (e.g., accounting measures of solvency). We also add that marker processes may be leading, lagging or coincident with respect to the parent process and their phase will be important in predictive inference for the parent process and its FHT.

## 9. DATA STRUCTURES FOR THRESHOLD REGRESSION MODELS

The data structures of threshold regression studies vary widely. To be specific, we look at the case in which longitudinal observations in calendar time are available for the parent process and the covariate vector. In this case, the data structure for a single individual can be summarized as follows:

(8)
Time points:
$0 = t_0 \leq t_1 \leq \cdots \leq t_m$,
Failure codes:
$f_0 = 0, f_1 = 0, \ldots, f_{m-1} = 0, f_m = 0$ or $1$,
Readings on parent process:
$x_0, x_1, \ldots, x_m$,
Covariate vectors:
$\mathbf{z}_0, \mathbf{z}_1, \ldots, \mathbf{z}_m$.

Each individual has observation vectors of the form $(t_j, f_j, x_j, \mathbf{z}_j)$, $j = 0, 1, \ldots, m$, where $t_0 = 0 \leq t_1 \leq \cdots \leq t_m$. Here $t_j$ is the time of the $j$th observation, $f_j$ is an indicator variable for whether the time $t_j$ is an FHT, $x_j$ is the state of the process at time $t_j$ and $\mathbf{z}_j$ is the covariate vector of the $j$th observation for the individual.

The data structures may have a variety of specialized features, as illustrated below.

1. The data sets usually consist of a sample of individuals, $i = 1, \ldots, n$, with individual parent processes $\{X_i(t)\}$ and boundary sets $\mathcal{B}^{(i)}$. The individual processes are often assumed to be mutually independent.
2. Where there are competing modes of failure, then the cause of failure $d$ will be recorded for each individual.
3. The final observation time $t_m$ for an individual is a random stopping time if $f_m = 1$. Thus, $t_m = S$ and $x_m = X(S)$ if $f_m = 1$. Here $X(S) \in \mathcal{B}$ is the threshold state realized by the individual at the FHT. If $f_m = 0$, then time $t_m$ is a right-censoring time for the FHT, that is, $t_m < S$. If $t_{m-1} < S \leq t_m$, then the survival time $S$ is interval censored.
4. The data are longitudinal if there is more than one reading available for some individuals, that is, if $m > 1$ for some individuals.
5. If the parent process is latent, then the data set will have no observations $x_j$, although there may still be readings on the covariate vectors $\mathbf{z}_j$.
6. If the data set consists only of a single time $t$ and failure indicator $f$ for each individual, then the data set constitutes censored survival data. With a baseline covariate vector $\mathbf{z}_0$ available, the data provide a basis for censored survival threshold regression.
7. Let $X(t_j)$ be abbreviated $X_j$ for any individual. The reading $x_j$ on the parent process, for $j < m$, is a realization of the conditional random variable $X_j | S > t_j$. The conditioning event is that the process has reached state $x_j$ at time $t_j$ without experiencing an FHT.
8. Where $\{X(t)\}$ is a Markov process (which is the most common type of model), we have for any individual that

$$P(X_j = x_j | x_{j-1}, \ldots, x_0, S > t_j)$$
$$= P(X_j = x_j | X_{j-1} = x_{j-1}, S > t_j)$$
$$\text{for } j < m.$$

In other words, the distribution of the next observation $X_j$ depends only on the value of the preceding observation $x_{j-1}$ and the fact that no FHT has yet occurred. The sample path by which $x_{j-1}$ was attained is immaterial.

Our discussion of data structures here has referred only to calendar time $t$ without reference to running time $r$. It also has not taken account of observations that may be available on relevant marker processes. The discussion in Sections 7 and 8, however, will make it clear what supplemental data are required when these model components are part of the threshold regression model.

## 10. PARAMETER ESTIMATION AND INFERENCE

In applications to date, parameter estimation for FHT models and threshold regression have been heavily dominated by maximum likelihood methods. The reason is that the probabilistic specification of the parent stochastic process in FHT models is usually explicit and, hence, likelihood expressions follow as a matter of course. The optimization required by this estimation method may employ a variety of computational techniques but gradient methods work very well. Extensions to Bayesian methods have been developed in some cases. For example, Pettit and Young (1999) and Shubina (2005a, 2005b) have embedded the Wiener diffusion FHT model in a Bayesian framework. Lee, DeGruttola and Schoenfeld (2000) have developed some predictive inference results for this case, in conjunction with a marker process. Nothing seems to stand in the way of developing nonparametric and semiparametric approaches for these models but these approaches have not yet been taken up in the literature.

*Case illustration.* To illustrate the nature of inference for one of the simple threshold regression settings, we now set up the sample log-likelihood function for censored inverse Gaussian regression for a medical application like that found in Lee et al. (2004). We consider a latent health status process defined on a running time scale $r$. We let the parent process be a Wiener diffusion process. The FHT for such a process follows an inverse Gaussian distribution. The inverse Gaussian distribution depends on the mean and variance parameters of the underlying Wiener process ($\mu$ and $\sigma^2$) and the initial health status level ($x_0$). We let $f(r|\mu,\sigma^2,x_0)$ and $F(r|\mu,\sigma^2,x_0)$ denote the probability density function (p.d.f.) and cumulative distribution function (c.d.f.) of the FHT distribution, both defined in terms of running time $r$. These functions have simple computational forms. For the case where the process begins at $x_0 > 0$ and the boundary is the zero level, the p.d.f. for the first hitting time is given by

$$f(r|\mu,\sigma^2,x_0) = \frac{x_0}{\sqrt{2\pi\sigma^2 r^3}} \exp\left[-\frac{(x_0+\mu r)^2}{2\sigma^2 r}\right] \quad (9)$$

for $-\infty < \mu < \infty, \sigma^2 > 0, x_0 > 0$.

If $\mu > 0$, then the FHT is not certain to occur and the p.d.f. is improper. Specifically, in this case, $P(X=\infty) = 1 - \exp(-2x_0\mu/\sigma^2)$. The c.d.f. corresponding to (9) is

$$F(r|\mu,\sigma^2,x_0) = \Phi\left[-\frac{(\mu r + x_0)}{\sqrt{\sigma^2 r}}\right] + \exp(-2x_0\mu/\sigma^2)\Phi\left[\frac{\mu r - x_0}{\sqrt{\sigma^2 r}}\right], \quad (10)$$

where $\Phi(\cdot)$ is the c.d.f. of the standard normal distribution.

The health status process is latent here and, hence, can be given an arbitrary measurement unit. Thus, one parameter may be fixed. We set the variance parameter $\sigma^2$ to unity. Both parameters $\mu$ and $x_0$ are linked to $k$ regression covariates that are represented by the row vector $\mathbf{z} = (1, z_1, \ldots, z_k)$. The leading 1 in $\mathbf{z}$ allows for a constant term in the regression relationship. An identity function of the form

$$\mu = \mathbf{z}\boldsymbol{\beta} = \beta_0 + \beta_1 z_1 + \cdots + \beta_k z_k$$

is used to link the parameter $\mu$ to the covariates and a logarithmic function

$$\ln(x_0) = \mathbf{z}\boldsymbol{\gamma} = \gamma_0 + \gamma_1 z_1 + \cdots + \gamma_k z_k$$

is used to link the parameter $x_0$ to the covariates. Here $\boldsymbol{\beta} = (\beta_0, \beta_1, \ldots, \beta_k)'$ and $\boldsymbol{\gamma} = (\gamma_0, \gamma_1, \ldots, \gamma_k)'$, where $\beta_0$ and $\gamma_0$ are regression constants. Parameters of the running time scale, such as $\boldsymbol{\alpha} = (\alpha_1, \ldots, \alpha_J)$ in (6), may also be linked to covariates using link functions of appropriate form.

We now denote $\mu$ and $x_0$ for subject $i$ by $\mu^{(i)}$ and $x_0^{(i)}$. We let $r^{(i)}$ denote the running time for subject $i$. Time $r^{(i)}$ is the running time at the moment of death for a dying subject and a right-censored running time for the moment of death for a surviving subject. Hence, each dying subject $i$ contributes probability density $f(r^{(i)}|\mu^{(i)}, x_0^{(i)})$ to the sample likelihood function, for $i = 1, \ldots, n_1$, and each surviving subject $i$ contributes survival probability $\overline{F}(r^{(i)}|\mu^{(i)}, x_0^{(i)}) = 1 - F(r^{(i)}|\mu^{(i)}, x_0^{(i)})$ to the sample likelihood function, for $i = n_1 + 1, \ldots, n_1 + n_0$. The sum $n = n_1 + n_0$ is the total number of subjects. The sample log-likelihood function to be maximized therefore has the form

$$\ln L(\boldsymbol{\alpha}, \boldsymbol{\beta}, \boldsymbol{\gamma}) = \sum_{i=1}^{n_1} \ln f(r^{(i)}|\mu^{(i)}, x_0^{(i)}) + \sum_{i=n_1+1}^{n_1+n_0} \ln \overline{F}(r^{(i)}|\mu^{(i)}, x_0^{(i)}). \quad (11)$$

Numerical gradient methods can be used to find maximum likelihood estimates for $\boldsymbol{\beta}$, $\boldsymbol{\gamma}$ and $\boldsymbol{\alpha}$.

## 11. THRESHOLD REGRESSION FOR LONGITUDINAL DATA ANALYSIS

Our discussion of data structures in Section 9 has anticipated that longitudinal data are gathered on the respective stochastic processes of individuals in some applications. Using our previous notation, we now let $\{A_j\}$ denote the *longitudinal observation process*, defined on the time points $t_j, j = 0, 1, \ldots$. If the individual survives beyond time $t_j$, then the failure code $f_j = 0$ and $A_j = \{S > t_j, x_j, \mathbf{z}_j\}$ for $j \leq m$. If the individual fails in the final interval $(t_{m-1}, t_m]$, then $f_m = 1$ and $A_m = \{S \in (t_{m-1}, t_m], x_m \in \mathcal{B}\}$. As defined earlier, $S$ is the stopping time for the longitudinal observation process. We note that $z_m$ is not defined when the individual has failed and, hence, is dropped from the expression for $A_m$. Moreover, the final reading $x_m$ for the parent process lies inside the boundary set $\mathcal{B}$ when the individual has failed.

Longitudinal data of this kind pose an interesting challenge for first-hitting-time models, as for most time-to-event models. Lu (1995) considers the problem for the basic Wiener model where longitudinal observations are made on the process $\{X(t)\}$ up to the hitting or censoring time, as the case may be. She formulates the likelihood function and computes exact maximum likelihood estimates. The methodology is somewhat intricate but manageable. Lee, DeGruttola and Schoenfeld (2000) consider the issue of modeling longitudinal data for a bivariate Wiener model representing a latent health status process and a correlated marker process. These authors mention an interesting approach to handling longitudinal data which they anticipated would be technically satisfactory and practical to implement. Their suggested approach, however, is not elaborated in their article, so we sketch one direction of development below but leave a full exploration of the approach as an open research question. We refer to this method as an *uncoupling* procedure because it effectively unlinks the longitudinal observations into a set of independent conditional observations.

With the preceding notation, the probability of observing the longitudinal data record of an individual can be expanded as a product of conditional probabilities as

$$(12) \quad \begin{aligned} &P(A_m, A_{m-1}, \ldots, A_1, A_0) \\ &= P(A_0) \prod_{j=1}^{m} P(A_j | A_{j-1}, \ldots, A_0). \end{aligned}$$

Now we come to the crucial assumption. If it can be assumed that $\{A_j, j = 0, 1, \ldots\}$ is a Markov process with initial state $A_0$, then (12) can be simplified as

$$(13) \quad \begin{aligned} &P(A_m, A_{m-1}, \ldots, A_1, A_0) \\ &= P(A_0) \prod_{j=1}^{m} P(A_j | A_{j-1}). \end{aligned}$$

In other words, the probability of observing $A_j$ depends only on its preceding state $A_{j-1}$ and not on the earlier history of the observation process. The explicit forms of the probability elements on the right-hand side of (13) are

$$(14) \quad \begin{aligned} &P(A_j | A_{j-1}) \\ &= P(S > t_j, x_j, \mathbf{z}_j | S > t_{j-1}, x_{j-1}, \mathbf{z}_{j-1}) \\ &\qquad \text{if } f_j = 0, j \leq m, \end{aligned}$$

$$(15) \quad \begin{aligned} &P(A_m | A_{m-1}) \\ &= P(S \in (t_{m-1}, t_m], x_m \in \mathcal{B} | \\ &\qquad S > t_{m-1}, x_{m-1}, \mathbf{z}_{m-1}) \quad \text{if } f_m = 1. \end{aligned}$$

If no observations are available on the parent process, then $x_j$ is dropped from the $A_j$ notation, giving $A_j = \{S > t_j, \mathbf{z}_j\}$ if $f_j = 0$, $j \leq m$, and $A_m = \{S \in (t_{m-1}, t_m]\}$ if $f_m = 1$. Again, invoking the Markov assumption for the observation process, (14) and (15) take the revised forms

$$(16) \quad \begin{aligned} &P(A_j | A_{j-1}) \\ &= P(S > t_j, \mathbf{z}_j | S > t_{j-1}, \mathbf{z}_{j-1}) \\ &\qquad \text{if } f_j = 0 \text{ for } j \leq m, \end{aligned}$$

$$(17) \quad \begin{aligned} &P(A_m | A_{m-1}) \\ &= P(S \in (t_{m-1}, t_m] | S > t_{m-1}, \mathbf{z}_{m-1}) \\ &\qquad \text{if } f_m = 1. \end{aligned}$$

Statement (13) is the theoretical justification for the *uncoupling procedure*. Neither this theoretical development nor issues of practical implementation of the procedure were taken up by Lee, DeGruttola and Schoenfeld (2000). As already noted, the procedure remains an open topic for future research.

## 12. MODEL VALIDATION, DIAGNOSTICS AND REMEDIES

Although procedures for model validation, diagnostics and remedies are not as well developed for threshold regression as for conventional regression models for survival data, a number of techniques have been proposed and applied successfully in earlier FHT investigations. For example, procedures are available for checking the assumptions of the TR regression model having a Wiener process and inverse Gaussian FHT, both with and without associated marker processes. Lee, DeGruttola and Schoenfeld (2000) present some procedures for this TR model and demonstrate the techniques using a medical case application. Lee and Whitmore (2002) present a larger suite of techniques for checking assumptions of this model and also discuss a number of remedies that might be used where assumptions do not hold. Lee et al. (2004) also discuss validation for an extension of this same model in which the calendar time scale is replaced by a job-exposure disease progression scale. One of the proposed validation procedures relies on the fact that the inverse Gaussian (IG) distribution is the first-stopping-time distribution of a Wiener process. Hence, comparisons can be made between Kaplan–Meier (KM) survival curves and the IG survival curves implied by the model (for different covariate subgroups). Applications with longitudinal observations on the parent process or marker measurements offer even more data for model validation. The previous work also points out the importance of having subject-matter specialists understand the model features and compare them with the fundamental physical processes at play. For example, the concept of an FHT is one feature whose mechanism is found frequently in nature, is easily understood by scientists and can be checked against their scientific understanding of the application context.

## 13. SOME OPEN RESEARCH PROBLEMS

Many interesting aspects of threshold regression require further study. We noted earlier that multicollinearity of parameter estimates can be a practical issue. It remains to be seen which parameterizations of threshold regression models tend to have relatively independent estimation errors. Multicollinearity within regression functions will tend to show itself in familiar ways and will likely be dealt with by conventional remedies.

The Cox proportional hazards regression model is widely used for survival data analysis. Threshold regression models do not generally possess the proportional hazards feature for different configurations of covariates. A useful research contribution would be made by comparing and contrasting the results of Cox regression and TR in the same context. Some public sets of survival data that are scientifically important and have a plausible FHT interpretation might very well be reanalyzed to see if the key research conclusions are materially affected when a TR model is used in place of a more conventional technique.

Both parent and marker processes may be subject to measurement error. For example, blood pressure is known to be measured with error. Whitmore (1995), for example, studied a Wiener diffusion FHT model with measurement error. The true state of a process is also often randomly masked. The incorporation of measurement or masking errors in TR models, where these extensions are motivated by significant applications, would represent a useful research extension.

The identification of individual marker processes and the construction of composite marker processes to track or mimic a latent parent process are challenging subjects that need further theoretical work and more experience with real applications. The challenge will be especially great where the marker processes and latent processes are from different classes of processes. A related open research issue concerns the investigation of whether markers are leading, lagging or coincident with the parent process.

Nonparametric, semiparametric and other robust estimation methods seem to have much to contribute to the successful application of threshold regression. Quasi- likelihood methods and generalized estimating equations may offer feasible approaches. As threshold regression estimation in a general setting involves parameter estimation for the boundary set, the parent process and the running time scale, it is conceivable that a blend of nonparametric and parametric methods may be effective in some applications. For example, nonparametric estimation of running time parameters might be combined with parametric estimation of the parent process.

Our discussion of the analysis of longitudinal data in the context of threshold regression has already pointed out that a full theoretical development and justification of the uncoupling method remains an open research issue. In the same vein, practical experience with this method or other methods for handling longitudinal data in threshold regression will be valuable contributions.

Much remains to be done on model validation and diagnostic techniques in the context of threshold regression. These tools are likely to be developed as threshold regression is applied in a broader range of practical cases. The earlier work on model validation has been largely restricted to the Wiener FHT model and thus extensions to other FHT models need attention. For example, comparisons of Kaplan–Meyer (KM) survival curves with fitted TR survival curves, both defined on running time scales, will require new methods that take account of the fact that the running time scale is itself fitted by a statistical model. As another example, TR models assume that particular functions link the model parameters to the regression covariates. Both the forms of the link functions and the adequacy of the regression functions must be validated. Whether the correct directing process has been chosen is also a feature that must be checked by model validation techniques. Although model validity is likely to be established by standard techniques (such as cross-validation), new techniques and modifications of conventional methods will surely be needed. In addition to using statistical methods for model verification, it is desirable to work closely with subject-matter specialists to ensure that the FHT models have realistic features and that the findings emerging from the analysis make practical sense.

The last sentence of the preceding paragraph hints at the largest open research question. Threshold regression will prove itself through beneficial practical application. With exploration of fresh application areas will come ideas for better methods and models for this new type of regression approach.

## ACKNOWLEDGMENTS

This research was supported in part by NIH Grants OH008649 and HL40619 (Lee) and by a research grant from the Natural Sciences and Engineering Research Council of Canada (Whitmore).## REFERENCES